\documentstyle[preprint,aps,epsfig]{revtex}

\tightenlines

\begin{document}
\title{
\vspace*{-2cm}
~ \hfill
\mbox{\small\rm IFIC-00-1013, FTUV-00-1013}\\
~\\
CHIRAL UNITARY APPROACH TO THE $K^-$ DEUTERON SCATTERING LENGTH}

\author{S.S. Kamalov$^a$\thanks{ Permanent address:
Laboratory of Theoretical Physics, JINR Dubna, 141980 Moscow
region, Russia.}, E. Oset$^a$ and A. Ramos$^b$ }

\address{$^a$ Departamento de F\'{\i}sica Te\'orica and IFIC, Centro Mixto
 Universidad de Valencia-CSIC, Institutos de Investigaci\'on de  Paterna,
Aptdo. Correos 22085, 46071  Valencia, Spain\\
$^b$ Departamento d'Estructura i Constituents de la Materia,
 Universitat de Barcelona, Diagonal~647, 08028 Barcelona, Spain}

\maketitle

\begin{abstract}
 Starting from a recent model where the $\bar{K} N$ amplitudes are evaluated
 from the chiral Lagrangians using a
 coupled channel unitary method, we evaluate here the scattering length for
 $K^-$ deuteron scattering. We find that the double scattering contribution is
 very large compared to the impulse approximation and that the charge exchange
 contribution of this rescattering is as large as the sequential $K^-$
 scattering on the two nucleons.  
 Higher order rescattering corrections are
 evaluated using coupled channels with $K^-$ and $\bar{K^0}$ 
 within the integral form of the fixed centre approximation to the Faddeev
 equations. The higher order corrections involving intermediate pions and 
 hyperons are found negligible.
\end{abstract}



\section{introduction}

     The low energy scattering of $K^-$ with deuterium has been the subject of
much study in the past \cite{landau,koltun} and it is one of the processes where
the impulse approximation is manifestly insufficient, the rescattering terms
being quite large. The input in all these studies is elementary amplitudes for
$\bar{K} N$  scattering which are either taken from experiment or evaluated
within theoretical models.  The theoretical models for $\bar{K} N$ are rather
involved since there are many coupled channels which have to be dealt with
consistently (concretely, 10 physical channels in the $K^- p$ channel, $K^-p$,
 $\bar{K}^0 n$, $\pi^+ \Sigma^-$, $\pi^- \Sigma^+$, $\pi^0 \Sigma^0$, $\pi^0
\Lambda$, $\eta \Sigma^0$, $\eta\Lambda$, $K^+ \Xi^-$, $K^0 \Xi^0$).
Theoretical studies with coupled channels were used in 
Refs.~\cite{dalitz,chand,martin,siegel} fitting the input to the data. In
Ref.~\cite{siegel} the
strength of the different transition potentials was determined from fits to the
data allowing only modifications of up to 50 percent from the SU(3) relations.

The introduction of chiral Lagrangians in the meson baryon sector
\cite{xpt} has allowed one to deal with this interaction from the
modern chiral perspective. Yet, a unitary treatment with coupled
channels is necessary in this case since perturbation theory
cannot be applied, among other reasons due to the presence of the
$\Lambda(1405)$ resonance below the $\bar{K} N$ threshold. Coupled
channel Lippmann Schwinger equations are used in 
Refs.~\cite{wolfram1,wolfram2} including the channels which are
physically open, and some terms from higher order chiral
Lagrangians are obtained from fits to experiment. In
Ref.~\cite{knangels} the $\eta \Sigma$ and $\eta \Lambda$ channels are
also included and a good description of the low energy data is
obtained, amongst them the properties of the $\Lambda(1405)$
resonance, which is generated dynamically from the lowest order
chiral Lagrangian with the coupled channel equations. In this
later case only the lowest order Lagrangian was used together with
a cut off which was the only free parameter of the theory. A
justification of the success of the method omitting the higher
order Lagrangians can be seen by comparing the similar success in the
meson meson sector of
the coupled channel equations using the lowest order Lagrangians
\cite{oller97} and the more refined Inverse Amplitude Method in
coupled channels \cite{oller99},
case which also includes the $O(p^4)$ Lagrangian (see also discussions
to this respect in Refs.~\cite{kektalk} and \cite{nsd}).

The case of $K^-$ deuteron scattering requires also the explicit
treatment of the coupled channels. The Faddeev equations rely
already on partial summations over the different channels which
lead to the ${\bar K}N$ $t$ matrix on each individual nucleon, but even
then the
explicit channels appear in the multiple collisions with two
different nucleons and the Faddeev equations can be generalized to
these channels \cite{gal,torres,mizutani}. In the present work we
follow these lines but we observe that the relevant channels in
the Faddeev equations are the $K^- N$ and $\bar{K}^0 N$ channels.
The channels involving $\pi \Sigma$, and other inelastic ones by
analogy, require at least three successive collisions on the
nucleons of the deuteron and provide a negligible contribution to
the deuteron scattering length. We simplify the Faddeev equations 
with the $\bar{K} N$ channels by means of the fixed centre 
approximation for the nucleons (FCA) and obtain the result already 
derived 
in \cite{chand}, although we do not indulge in the isospin symmetry
assumed there. The results obtained in this formalism 
improve on those reported in \cite{barret,deloff} using the FCA 
approach and including only the elastic $K^-$ collisions. 
Nevertheless, the main novelties of the present results come from the 
use of the new 
elementary amplitudes obtained in the recent chiral approach 
to the $\bar{K} N$ interaction.

The measurement of the $K^-$ deuteron scattering length can thus
provide information on some of the $\bar{K} N$ elementary
 amplitudes provided the others
are already known. Assuming isospin symmetry, the knowledge of the
$K^- p$ scattering amplitude allows one to obtain the $K^- n$
scattering length using the Faddeev formula. However, one of the
findings of this work and the one of Ref.~\cite{knangels} is that
isospin symmetry is not very accurate for energies close to
threshold so one has to admit certain uncertainties when
extracting the elementary amplitudes from the deuteron
scattering data. In any case the deuteron results will provide
extra checks of accuracy of the modern chiral theories used for
the $\bar{K} N $ interaction.

Our treatment involves only the evaluation of the strong
interaction scattering length.  Coulomb corrections to the Deser
formula \cite{deser} to extract the scattering length from the
measurement of the width and shift of the 1s level of the $K^-$
deuteron atom planned at Frascati~\cite{guaraldo} have been worked
out in Ref.~\cite{barret}.

\section{$K^-N$ scattering lengths}

As mentioned in the introduction, the dynamics of ${\bar
K}N$ scattering at low energies is dominated by the
presence of the $\Lambda(1405)$ resonance and needs to be
described by non-perturbative methods. 
In this section we review the approach followed in Ref.~\cite{knangels}
and present
the results for the scattering lengths of the elementary ${\bar K}N$
reactions needed in the calculation of the $K^- d$
scattering amplitude, namely
$K^- p \to K^- p$, $K^- n \to K^- n$, $\bar{K}^0 n \to \bar{K}^0 n$ and 
$K^- p \to \bar{K}^0 n$.

The starting point is the lowest-order
chiral Lagrangian coupling mesons and baryons, which in the case of
meson-baryon
transition amplitudes reduces to
\begin{equation}
L_1^{(B)} = \langle \bar{B} i \gamma^{\mu} \frac{1}{4 f^2}
[(\Phi \partial_{\mu} \Phi - \partial_{\mu} \Phi \Phi) B
- B (\Phi \partial_{\mu} \Phi - \partial_{\mu} \Phi \Phi)]
\rangle     \ ,
\label{eq:chiral}
\end{equation}
where 
$\Phi$ and $B$ denote the octets of pseudoscalar mesons and $1/2^+$
baryons,
respectively, and
the symbol $\langle \rangle$ stands for the trace of SU(3) matrices.

{}From the Lagrangian of Eq. (\ref{eq:chiral}) one derives all possible
transition amplitudes between the different meson-baryon states of a given
charge and strangeness that
can be
built from the meson and baryon octets. 
There are ten such channels for $K^- p$ scattering, namely
$K^-p$, $\bar{K}^0 n$, $\pi^0
\Lambda$, $\pi^0 \Sigma^0$,
$\pi^+ \Sigma^-$, $\pi^- \Sigma^+$, $\eta \Lambda$, $\eta
\Sigma^0$,
$K^+ \Xi^-$ and $K^0 \Xi^0$, and six in the case of $K^- n$ scattering,
namely $K^-n$, $\pi^0\Sigma^-$,
 $\pi^- \Sigma^0$, $\pi^- \Lambda$, $\eta
\Sigma^-$ and
$K^0 \Xi^-$. At low energies all the possible 
amplitudes reduce to the form
\begin{equation}
V_{i j} = - C_{i j} \frac{1}{4 f^2} (k_j^0 + k_i^0) \ ,
\label{eq:v}
\end{equation}
where $k_j^0, k_i^0$ are the initial, final energies of the
mesons and
the explicit values of the coefficients $C_{ij}$ can be
found in Ref.~\cite{knangels}.

Using average masses for each particle multiplet, it is also possible to
work in isospin formalism. Making the appropriate basis transformation,
the transition coefficients of good isospin, $D_{ij}$ for $I=0$ channels
(${\bar K}N$, $\pi\Sigma$, $\eta\Lambda$, $K\Xi$) and
$F_{ij}$ for $I=1$ ones (${\bar K}N$, $\pi\Sigma$, $\pi\Lambda$,
$\eta\Sigma$, $K\Xi$), can be easily
derived from the $C_{ij}$ coefficients involving  $K^-p$ and related
channels and are also given in Ref.~\cite{knangels}. 

The lowest-order amplitudes of Eq.~(\ref{eq:v}) are then inserted in 
a coupled-channel Bethe-Salpeter
equation 
\begin{equation}
t_{i j} = V_{i j} + V_{i l} \; G_l \; t_{l j} \ ,
\end{equation}
from where one
extracts the elastic and transition scattering amplitudes. The indices
$i,l,j$ run over all possible meson-baryon channels and $G_l$ is the 
loop function containing the propagators of the meson and baryon in the
intermediate states.
Although in
the former equation the last term on the right hand side involves in
principle the off-shell dependence of the amplitudes, 
the simple form of $V_{ij}$ in Eq. (\ref{eq:v}) allows to reabsorb the
off-shell pieces of the amplitude into
renormalization of
the coupling constant $f$, as shown in Ref.~\cite{knangels}. Therefore,
the $V$ and $t$ amplitudes simply factorize
on-shell out of the loop integral and the problem reduces to one of
solving a coupled set of algebraic equations, with $G_l$ given by
\begin{eqnarray}
G_{l}(\sqrt{s}) &=& i \, \int \frac{d^4 q}{(2 \pi)^4} \,
\frac{M_l}{E_l
(-\vec{q}\,)} \,
\frac{1}{\sqrt{s} - q^0 - E_l (-\vec{q}\,) + i \epsilon} \,
\frac{1}{q^2 - m^2_l + i \epsilon} \nonumber \\
&=& \int_{\mid {\vec q} \mid < q_{\rm max}} \, \frac{d^3 q}{(2
\pi)^3} \,
\frac{1}{2 \omega_l (\vec q\,)}
\,
\frac{M_l}{E_l (-\vec{q}\,)} \,
\frac{1}{\sqrt{s}- \omega_l (\vec{q}\,) - E_l (-\vec{q}\,) + i
\epsilon} \ ,
\label{eq:gprop}
\end{eqnarray}
where $M_l, E_l$ and $m_l$ stand, respectively, for the
baryon mass, the baryon energy and the meson mass in the intermediate
state, and $\sqrt{s}$ is the total energy in the center-of-mass (CM)
frame.

The approach of Ref.~\cite{knangels} and summarized here depends on one
parameter, the
loop regularization
cut-off, $q_{\rm max}$. Using the particle basis, 
a value of $630$ MeV was 
adjusted to reproduce the $K^- p$ scattering branching ratios
at threshold. At the same time the weak decay constant
was
slightly modified to $f=1.15
f_\pi$, a value lying in between the empirical pion and kaon weak decay
constants, in order to optimize
the position of the $\Lambda(1405)$ resonance.
The scattering cross
sections, which were not used in the fit, were shown to be in good
agreement with the low energy data.

The scattering lengths are obtained from the amplitudes
$t_{ij}$ through
\begin{equation}
a_{ij} = -\frac{1}{4\pi}\frac{m_N}{\sqrt{s}}t_{ij} \ ,
\end{equation}
and the relevant ones for the study of $K^- d$ scattering are shown in
Table \ref{tab:1}.  
We also give there the results for the scattering lengths from
the isospin formalism, obtained 
by first solving the Bethe-Salpeter equation for the
various isospin channels and then transforming the isospin amplitudes  
back to the particle basis.

\section{Multiple scattering series }

It is well known that the impulse approximation fails to describe
the $K^-$ deuteron scattering length. Furthermore even the
contribution of a few terms of the multiple scattering series does
not give accurate values for the scattering length. This indicates
that the multiple scattering series does not converge rapidly for
$K^-d$ elastic scattering at low energies. Therefore, in this case
more sophisticated approaches based on the solution of the Faddeev
equations are required.

On other hand, there are many difficulties in the solution of the
Faddeev equations for the $K^-d$ elastic scattering. The first one
is related to the coupling to many inelastic channels with
$\Sigma,\,\Lambda$, and $\Xi$ baryons which make the solution of
the problem technically difficult. The second problem is related
to the isospin symmetry which is often used in the solution of the
Faddeev equations. Indeed, as seen from the considerations in the
former section, isospin for $K^-N$ scattering is a good
quantum number only with accuracy of 20\%. In such a situation the
use of the physical channels for the coupled equations becomes
more realistic than using an isospin formalism. This of course
increases the number of the coupled channels making the numerical
procedure more complicated.

Thus getting an unambiguous information about $K^-N$ scattering
lengths from elastic $K^-d$ scattering  becomes a difficult task.
However, below we present one theoretical scheme which should
considerably facilitate the solution of this problem.

\subsection{Single scattering (impulse) approximation}

We will start our considerations with the well known results of
the {\it impulse approximation} where only contributions from one
(single) kaon scattering are taken into account. In this case, we
can get the following expression for the $s$-wave $K^-d$
scattering $t$-matrix ($T_{Kd}$) in terms of the elementary
$s$-wave $t$-matrices which describe $K^-N$ scattering on the
proton ($t_p$) and neutron ($t_n$):
\begin{equation}
 T_{Kd}(k',k)\, =\, [ t_p(k',k) + t_n(k',k) ]\,F_d(Q)\,,
\label{sk1}
\end{equation}
where ${\bf Q}=({\bf k}'-{\bf k}) /2$ is the momentum transfer
with initial and final kaon momentum ${\bf k}$ and ${\bf k}'$,
respectively. $F_d(Q)$ is the elastic deuteron form factor
\begin{equation}
 F_d(Q) = \int e^{-i{\bf Q}\cdot{\bf r}}\,
 \mid \phi_d({\bf r})\mid^2\, d{\bf r}
\label{sk2}
\end{equation}
normalized to unity at $Q=0$. Therefore $ \mid \phi_d({\bf
r})\mid^2= \mid u(r)\mid^2+ \mid w(r)\mid^2$, where $u(r)$ and
$w(r)$ are the $S$- and $D$-components of the deuteron wave
functions taken from Ref.~\cite{Lacombe}.

For the low energy limit, when $ {\bf k,\,k'} \rightarrow 0$ ,
taking into account the relations between $t$-matrices and
scattering lengths (or amplitudes)
\begin{equation}
 T_{Kd}= -\frac{4\pi (m_K + M_d)}{M_d} A_{Kd},\qquad
 t_{p,n}= -\frac{4\pi (m_K + m_N)}{m_N} a_{p,n}
\label{sk3}
\end{equation}
we obtain the following simple expression for the $K^-d$
scattering length in the impulse approximation
\begin{equation}
 A_{Kd}^{IA}=\frac{M_d}{m_K + M_d}\left(1+\frac{m_K}{m_N}\right)\,
 ( a_p + a_n)=(-0.26+i\,1.87)\,{\rm fm}\,,
\label{sk4}
\end{equation}
where $M_d$ and $m_N$ are the deuteron and nucleon masses,
respectively, and $m_K$ is the kaon mass. The numerical value was
obtained using the elementary amplitudes $a_p$ and $a_n$ in the
physical basis at $W=1432.6$ MeV, which corresponds to having the proton
and neutron on shell at rest. The amplitudes can be seen in
Table~\ref{tab:1}.

Note that in general within the impulse approximation the effects
from the motion of the nucleons has to be taken into account in
the evaluation of the elementary $t$-matrix. However, numerous
investigations \cite{Mach,TRD,EGK,Boffi} show that the
substitution for the nucleon momentum ${\bf p}_N \rightarrow {\bf
p}_{eff.}=-({\bf k}-{\bf k'})/2$ is a very good approximation. In
the case of $S$-shell nuclei such approximation is even exact for
the linear terms in ${\bf p}_N$. Therefore, we expect that in the
limit ${\bf k}\rightarrow 0$ the static approximation, ${\bf
p}_N=0$, is reliable.

\subsection{ Double scattering contribution}

\begin{figure}[h]
\centerline{\psfig{file=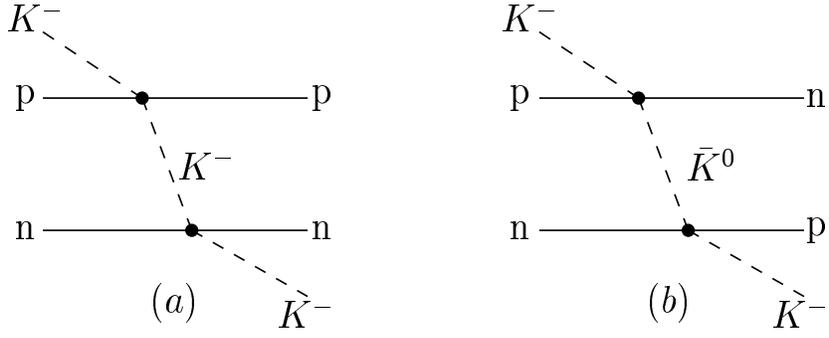,width=11cm,silent=}}
\vspace{0.5cm} \caption{ Graphical illustration of the double
scattering contributions } \label{fig1_kor}
\end{figure}

The first correction to the impulse approximation is related to
the contributions coming from the diagrams depicted in Fig.
\ref{fig1_kor}. We evaluate them using Feynman diagram rules. Then
for the $S$-matrix we get
\begin{eqnarray}
 S^{(2)}_{Kd} & = & \int \int d^4x\,d^4x' \,
 \frac{1}{\sqrt{2V\omega_K}} \, \frac{1}{\sqrt{2V\omega_{K'}}} \,
 \varphi_p^*(x)\, \varphi_n^*(x')\, e^{-iKx}\,e^{iK'x'}\,
 \varphi_p(x)\, \varphi_n(x') \nonumber
 \\ & \times &  \int \frac{d^4 q}{(2\pi)^4}\,\frac{ i\,e^{iq(x-x')}}
 {q^2-m_K^2 + i\epsilon}\, (-i t_p)\,(-i t_n)\,,
\label{sk5}
\end{eqnarray}
where $K=(\omega_K,{\bf k})$ and  $K'=(\omega'_K,{\bf k'})$ are
kaon 4-momenta in the initial and final states and
$\varphi_{p(n)}$ is the proton (neutron) wave function normalized
to unity. The plane waves are normalized to unity in the volume
$V$. The space part of the deuteron wave function in Eq.
(\ref{sk5}) can be written in terms of CM and relative
coordinates, i.e.
$\varphi_p({\bf x})\, \varphi_n({\bf x}')=\frac{1}{\sqrt{V}}
e^{i{\bf K}_d\cdot {\bf R}}\varphi_d({\bf r}).
$
Then the $S$-matrix can be related to the kaon-deuteron scattering
$T$-matrix in the following way
\begin{equation}
S^{(2)}_{Kd}\,=\,1\,-\,i\frac{(2\pi)^4\,M_d}{V^2\sqrt{2\omega_K\,
2\omega'_K\,E_d\,E'_d}}\,\delta(K+K_d-K'-K_d)\, T^{(2)}_{Kd}\,,
\label{sk6}
\end{equation}
where $E_d$ and $E'_d$ are the total energies of the deuteron with
momentum ${\bf K}_d$ and ${\bf K}'_d$ in the initial and final
states, respectively.

In the CM frame and low energy limit ${\bf k,\,k'}\rightarrow 0$
we obtain the following expressions and numerical values
for the contribution from diagrams (a) and (b) of Fig.
\ref{fig1_kor}
\begin{mathletters}
\begin{equation}
 A_{Kd}^{(2,a)}=\frac{M_d}{m_K + M_d}\left(1+\frac{m_K}{m_N}\right)^2\,
 2\,a_p\, a_n\,\left<\frac{1}{r}\right>=(-1.70+i\,0.07)\,{\rm fm}\,,
\label{sk7a}
\end{equation}
\begin{equation}
 A_{Kd}^{(2,b)}=-\frac{M_d}{m_K + M_d}\left(1+\frac{m_K}{m_N}\right)^2\,
 \,a_x^2\,\left<\frac{1}{r}\right>=(-0.77+i\,0.95)\,{\rm fm}\,,
\label{sk7b}
\end{equation}
\end{mathletters}
where
\begin{equation}
 \left<\frac{1}{r}\right>=\frac{2}{\pi}\,\int_{0}^{\infty}\,F_d(q)\,dq
 = \int d{\bf r} \mid \varphi_d(r)\mid^2\,\frac{1}{r}=0.449 \,{\rm
 fm}^{-1}
\label{sk8}
\end{equation}
and $a_x$ is the scattering length (or amplitude) for the kaon
charge exchange reaction $K^- p \rightarrow {\bar K}^0 n$. Thus we
can see that the contribution to the real part of $A_{Kd}$ from
the double scattering is larger than that from the impulse
approximation. This is mainly due to the cancellation of the
proton and neutron contributions from the single scattering.

\subsection{Triple scattering and coupling with the $\Sigma\pi$
channel}

\begin{figure}[h]
\centerline{\psfig{file=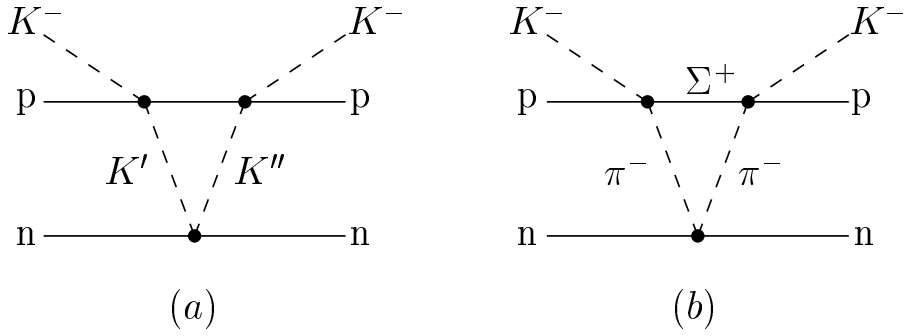,width=12cm,silent=}}
\vspace{0.5cm} \caption{ Graphical illustration of the triple
scattering contributions} \label{fig2_kor}
\end{figure}

 Now let us estimate the contribution from the diagrams
depicted in Fig. \ref{fig2_kor}a. The corresponding $S$-matrix is
\begin{eqnarray}
 S^{(3)}_{Kd} & = & \int\int \int d^4x\,d^4x'\,d^4x'' \,
 \frac{1}{\sqrt{2V\omega_K}} \, \frac{1}{\sqrt{2V\omega_{K'}}} \,
 \varphi_p^*(x'')\, \varphi_n^*(x')\, e^{-iKx}\,e^{iK'x''}\,
 \varphi_p(x)\, \varphi_n(x') \nonumber
 \\ & \times &  \int \frac{d^4 q}{(2\pi)^4}\,\frac{ i\,e^{iq(x-x')}}
 {q^2-m_K^2 + i\epsilon}\, \int \frac{d^4 q'}{(2\pi)^4}\,\frac{
 i\,e^{iq'(x'-x'')}} {q'^2-m_K^2 + i\epsilon}\, (-i t_p)\,(-i
 t_n)\,(-i t_p)\nonumber \\ & \times &
 \int \frac{d^4 p}{(2\pi)^4}\,\frac{ i\,e^{ip(x-x'')}}
{p^0-E({\bf p}) + i\epsilon}\,.
\label{sk9}
\end{eqnarray}
First let us evaluate the contribution from triple kaon scattering
including also charge exchange processes. In contrast to the case
of the double scattering considered above, now we have two mesons
and one baryon propagators. As a first step we evaluate exactly
the energy variable integration. After this we make the assumption
of heavy baryons, i.e. $E({\bf p})$ in the baryon propagator is
replaced by the baryon mass. Then the integration over the three momentum
of the baryon ${\bf p}$ gives rise to a $\delta^3({\bf x}-{\bf
x}'')$ function which brings together the ${\bf x}$ and ${\bf
x''}$ coordinates. Formally this is equivalent to the so called
{\it fixed centre approximation} (FCA) often used in the literature 
\cite{chand,barret,deloff}.
Using this approximation and taking into account that at low
energies the on-shell kaon energy $\omega_K\rightarrow m_K$
\begin{equation}
 \int\,\frac{d{\bf q}}{(2\pi)^3}\, \frac{e^{-i{\bf q}\cdot {\bf
 r}}} {\omega_K^2-m_K^2-{\bf q}^2+i\epsilon} \rightarrow
 -\,\frac{1}{4\pi r} \ ,
\label{sk10}
\end{equation}
we obtain the following expression and numerical value for the
triple kaon scattering contribution
\begin{eqnarray}
\label{sk11}
 A_{Kd}^{(3)}&=&\frac{M_d}{m_K + M_d}\left(1+\frac{m_K}{m_N}\right)^3\,
[a_p a_n (a_p +
 a_n)-a_x^2(2a_n-a_n^0)]\left<\frac{1}{r^2}\right> \nonumber \\ & =
 & (-1.12 +i\,0.07)\,{\rm fm}\,,
\end{eqnarray}
where $<1/r^2>=\int d{\bf r} \mid
\varphi_d(r)\mid^2/{r^2}=0.289\,{\rm fm}^{-2}$ and the amplitude
$a_n^0$ describes elastic scattering of the $\bar{K}^0$ meson on
the neutron. Note, by comparing Eq. (\ref{sk11}) with Eqs.
(\ref{sk4}), (\ref{sk7a}), (\ref{sk7b}) that the convergence of the
multiple scattering series is rather poor.

To estimate the contribution which would come from the coupling
with the $\Sigma \pi$ channel we, as an example, consider the
$\Sigma^+ \pi^-$ channel. The corresponding diagram is depicted in
Fig. \ref{fig2_kor}b. For its evaluation we will use again the
FCA approach. The only difference is in the
treatment of the pion propagator. Now  the on-shell pion energy in
the intermediate state is $\omega_{\pi}=220$ MeV and it is well
known that in this region the $p$-wave contribution related with
the excitation of the $\Delta(1232)$ resonance dominates,
especially in the $\pi^-n$ elastic channel. Therefore, for the
estimation we shall consider the contribution only from the
$\Delta$ resonance taking $t_{\pi^-n}\approx
t_{\pi^-n}^{\Delta}\,{\bf q}\cdot{\bf q}'$. In the description of
the pion propagation we will use the so called K-matrix (or
on-shell) approximation, i.e
\begin{equation}
\frac{1}{\omega^2-m_{\pi}^2-{\bf q}^2+i\epsilon}\rightarrow
-i\pi\,\delta(\omega^2_{\pi}-m_{\pi}^2-{\bf q}^2)\,.
\label{sk12}
\end{equation}
For these energetic pions this approximation allows one to take
into account the largest part of the pion rescattering
contribution. The final expression obtained in this way for the
contribution to the scattering length is the following
\begin{equation}
 A_{Kd}^{(\Sigma)}=-\frac{M_d}{m_K + M_d}\left(1+\frac{m_K}{m_N}\right)^2\,
 \left(1+\frac{m_{\pi}}{m_N}\right)
 \,a_{\Sigma}^2\,a_{\pi^-n}^{(p)}\, \left< q^2\,j_1^2\right>\,,
\label{sk13}
\end{equation}
where $q=174$ MeV is the on-shell pion momenta, $j_1(z)$ is the
spherical Bessel function and
\begin{equation}
 \left< q^2\,j_1^2\right>\, =q^2\,\int d{\bf r} \mid
 \varphi_d(r)\mid^2 \,j_1^2(q r) = 0.083\,{\rm fm}^{-2}\,.
\label{sk14}
\end{equation}
The value of the $p$-wave part of the $\pi^-n$ elastic scattering
amplitude is $a_{\pi^-n}^{(p)}=0.50+i0.09$ fm which corresponds to
the $\Delta(1232)$ contribution at $\omega_{\pi}=220$ MeV. Now if
we take for the amplitude of the $K^-p\rightarrow\Sigma^+\pi^-$
reaction the value $a_{\Sigma}=-0.39+i0.04$ fm from
Ref.~\cite{knangels}, we obtain that
$A_{Kd}^{(\Sigma)}=-0.015+i0.000$ fm which is only 1\% of the
contribution from the triple scattering given by Eq. (\ref{sk11}).
This estimation allows with a good accuracy to neglect the
contributions from the coupling with inelastic channels, but we
must keep the coupling with the kaon charge exchange channel. For
this purpose in the next section we derive a formalism based on
the FCA approach to the Faddeev equations.

\section{Solution of the Faddeev equations in the FCA approach}

As we have demonstrated in the previous section the convergence of the
multiple scattering series is very poor. Therefore, we can not use
the iteration procedure to calculate the $K^-d$ scattering length.
For this purpose we follow a more general scheme based on the
solution of the Faddeev equations. Note that since isospin is not a good
quantum number, we will write these equations in the physical
basis and present the elastic scattering $T$-matrix as a sum of the
two Faddeev partitions
\begin{equation}
    T_{Kd}\,=\,T_p\,+\,T_n\,,
\label{sk15}
\end{equation}
where $T_p$ and $T_n$ describe the interaction of the $K^-$-mesons
with the deuteron starting with a first collision on a proton and
a neutron, respectively.

\begin{figure}[h]
\centerline{\psfig{file=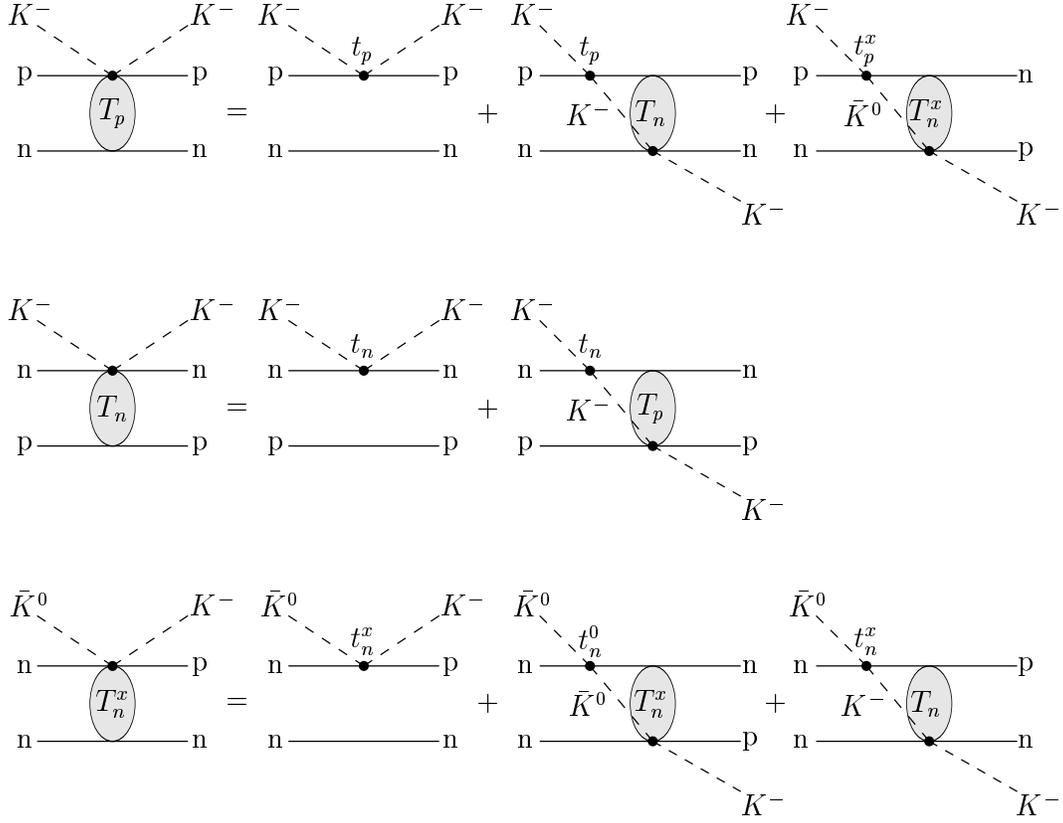,width=14cm,silent=}}
\vspace{0.5cm} \caption{ Graphical illustration of the Faddeev
partitions in kaon-deuteron scattering} \label{fig3_kor}
\end{figure}

Graphically these interactions are illustrated in Fig.
\ref{fig3_kor} and they satisfy the following system of integral
equations
\begin{eqnarray}
\label{sk16}
 T_p & = & t_p + t_p\,G_0\,T_n + t_p^x\,G_0\,T_n^x \,,\\
 T_n & = & t_n + t_n\,G_0\,T_p \,,\nonumber \\
 T_n^x & = & t_n^x + t_n^0\,G_0\,T_n^x +
 t_n^x\,G_0\,T_n\,,\nonumber
\end{eqnarray}
where $G_0$ is the free kaon propagator and $t_p$ and $t_n$ are
the $t$-matrices for $K^-p$ and $K^-n$ elastic scattering,
respectively. Note that for the proton partition, $T_p$, we have
also a contribution from the charge exchange channel with
elementary $t$-matrix, $t_p^x$, and the third Faddeev partition,
$T_n^x$, which describes the ${\bar K}^0 nn\rightarrow K^-pn $
transition including multiple rescattering in the intermediate
inelastic states. Through this term the coupling with the break-up
channel is realized and it is expressed via the additional
elementary charge exchange, $t_n^x$, and elastic ${\bar K}^0 n$
scattering, $t_n^0$, matrices.

The equations (\ref{sk16}) are a set of operator equations. On the
other hand, the final expression for the scattering length appears
as an expectation value of the scattering operator with the
deuteron ground state, i.e.
\begin{equation}
 A_{Kd}\,=\frac{M_d}{m_K+M_d}\int\,d{\bf r}\,\mid \varphi_d({\bf r})\mid^2\,
 {\hat A}_{Kd}(r)\,,\qquad {\hat A}_{Kd}(r)={\hat A}_p(r)+{\hat A}_n(r)
\label{sk17}
\end{equation}
Indeed, the analytical expressions for the amplitudes ${\hat
A}_p(r)$ and ${\hat A}_n(r)$ in Eq. (\ref{sk17}) can be determined
by the solution of the Faddeev-like equations (\ref{sk16}). For that,
let us apply recipes which we have found in the calculations of
the multiple scattering series. First, following Eq. (\ref{sk10})
the integral over the kaon propagator $G_0$ is replaced by
$-1/4\pi r$. Second, using the relations (\ref{sk3}), all the
elementary matrices $t_p,\,t_n,\,t_p^x=t_n^x$ and $t_n^0$ are
replaced by their threshold values of the corresponding scattering
lengths $a_p,\,a_n,\,a_x$ and $a_n^0$, respectively, up to a
factor. Then we get the following system of equations for the
amplitudes ${\hat A}_p(r)$ and ${\hat A}_n(r)$:
\begin{eqnarray}
\label{sk18}
 {\hat A}_p(r) & = & {\tilde a}_p + {\tilde a}_p\frac{1}{r}\,{\hat
 A}_n(r) - {\tilde a}_x\frac{1}{r}\,{\hat A}_n^x(r)\,,\\
 {\hat A}_n(r) & = & {\tilde a}_n + {\tilde a}_n\frac{1}{r}\,{\hat
 A}_p(r)\,, \nonumber \\
 {\hat A}_n^x(r) & = & {\tilde a}_x - {\tilde a}_n^0\frac{1}{r}\,
 {\hat A}_n^x(r) + {\tilde a}_x\frac{1}{r}\,{\hat
 A}_n(r)\,,\nonumber
\end{eqnarray}
where ${\tilde a} = a\,(1+m_K/m_N)$. Note that there is a minus
sign in the terms which lead to $np$ configuration in the final
state due to the fact that this configuration appears with minus
sign in the isospin zero wave function of the deuteron,
$(pn-np)/\sqrt{2}$.

After the solution of the system of equations (\ref{sk18}) the
amplitude ${\hat A}_{Kd}$ can be written in an analytic form
\begin{equation}
 {\hat A}_{Kd}(r)=\frac{{\tilde a}_p + {\tilde a}_n +(2{\tilde
 a}_p{\tilde a}_n-b_x^2)/r - 2b_x^2{\tilde a}_n/r^2}{1-{\tilde a}_p
 {\tilde a}_n/r^2 + b_x^2{\tilde a}_n/r^3}\,,
\label{sk19}
\end{equation}
where $b_x={\tilde a}_x/\sqrt{1+{\tilde a}_n^0/r}$ is the charge
exchange amplitude renormalized due to the ${\bar K}^0 n$
rescattering. Eq. (24) is then equivalent to the result obtained in \cite{chand},
 with the difference that isospin symmetry was assumed there, while here 
 $b_x$ and $a_n^0$ are not related to $a_p$ and $a_n$ by this symmetry.
Our result also contains the recoil factors $M_d/(m_K + M_d)$ and
$1+m_K/m_N$. 
 Eq. (24), together with Eq. (22) which produces the weigthed average
of Eq. (24) with the deuteron wave
function, is what is called FCA-integ in
 \cite{deloff}, although the charge exchange terms as well as recoil
factors are omitted there. Another
 option of the FCA quoted there, called FCA-aver, is the use in Eq. (24)
of an average value of $r$, for which the root mean square
radius of the deuteron is taken.

 If we keep only terms of order $1/r$ in the
solution of eq. (24) we get
\begin{equation}
 {\hat A}_{Kd}^{(1)}(r)={\tilde a}_p + {\tilde a}_n
 +(2{\tilde a}_p{\tilde a}_n-{\tilde a}_x^2)\frac{1}{r}
\label{sk20}
\end{equation}
which brings us to the {\it (IA + double scattering)} results in the
multiple scattering approach [see Eqs.
(\ref{sk4}), (\ref{sk7a}), (\ref{sk7b})]. In a similar way, by expanding
up to order $(1/r)^2$ we can easily obtain our previous results
for triple scattering.

 Eq. (25) is an approximation to eq. (24) valid only for large values of $r$. 
 In fig. 4 we compare the two results which show the importance of multiple 
 scattering in providing the right contribution at short and intermediate
 distances.
 
In Table \ref{tab:2} we collect our final results obtained using the
elementary amplitudes in the physical and isospin basis. Here we
again demonstrate the poor convergence of the multiple scattering
series. The calculations are done at $W=1432.6$ MeV which corresponds to
having a
neutron and a proton on shell at rest. One may wonder how the binding energy of
the deuteron affects these results. If one considers a simple shift of $W$ 
in the argument of the elementary amplitudes to
account for the binding, this leads
to corrections of the order of 20\%. 
However, we have checked that if this binding is considered
selfonsistently, i.e. both in the external energy and also as a nucleon
mass shift in the intermediate nucleon propagators, the corrections
obtained are negligible.

We can see that the results obtained here using elementary
scattering amplitudes that relied upon isospin symmetry differ
somewhat from those obtained using the elementary amplitudes
calculated with the physical basis. Particularly, the imaginary
parts of the scattering length differ by about 30\%. In addition,
note that even if we take the $a_n$ and $a_p$ amplitudes from the
physical basis and for the others we use the isotopic relations
$a_x=a_p-a_n$ and $a_n^0=a_p$, we get $A_{Kd}=-2.10+i\,1.90$ fm.
Comparison with the full results, $-1.61+i\,1.91$ fm, obtained within
the physical basis further demonstrates the consequences of the
isospin violation effects for $K^-$-deuteron scattering.

 With respect to the approach of Ref.~\cite{barret}, which uses a similar
method
to ours, we have included the charge exchange channels. We can see
from Eqs. (\ref{sk7a}),(\ref{sk7b}) that the charge exchange double
scattering
is rather important and this is also the case when the full
multiple scattering series is summed, as one can see in Table~\ref{tab:2}.
There we show the results obtained from the multiple scattering
series neglecting the charge exchange contribution ($b_x=0$) in
Eq. (\ref{sk19}), which we call ``only el.resc.".  The
``charge exch." results in the table denote the changes induced
by the term $b_x$, i.e.
$A_{Kd}$(charge exch.)$= A_{Kd}$(total)$-A_{Kd}$(only el.resc.).
Our result for the scattering
length,
$A_{Kd}=-1.61+i\,1.91$ fm, has larger strength for both the real
and imaginary parts that those found in Ref.~\cite{barret}, around
$-0.7 + i\,1.2$ fm or those of Ref.~\cite{gal}, $-1.47 + i\, 1.08$ fm.

 It is also worth comparing our results with those of Ref.~\cite{deloff},
where the author investigates the differences between the FCA and a
genuine Faddeev calculation, which solves the problem of
 the interaction of the three particles involved rather than using the partial 
 solution of the nucleon nucleon problem in terms of the deuteron wave 
 function. The comparison in Ref.~\cite{deloff} is done solving the
Faddeev equations and the FCA with only the $K^-$ channel, but one of the
conclusions of the
 author is that the coupled channel solution is necessary for accurate results.
   It is also concluded there that the FCA approach in the integral form
is a rather 
 good approximation to the Faddeev equations. Yet, the result quoted as best in
 this paper from the multichannel approach of Ref.~\cite{torres} is 
 $A_{Kd}=-1.34+i\, 1.04$ fm while ours is $A_{Kd}=-1.61 +i\, 1.91$ fm.
There seems
 to be large discrepancies, particularly in the imaginary part.  Actually,  
 the 
 first of these two amplitudes is evaluated with input that leads to different
 elementary scattering amplitudes than the ones we have used. These
amplitudes at threshold are
 quoted in Table II of \cite{deloff} and we have included them in the
last column of our Table~\ref{tab:1}, where we can see that they are
rather
different than those used here.
 For the purpose of comparison we recalculated
the $A_{Kd}$ amplitude in our approach, using the elementary
scattering lengths of
Ref.~\cite{deloff} together with isospin symmetry to generate $a_x$ and
$a_n^0$  from $a_p$ and $a_n$, namely  $a_x=a_p - a_n$ and $a_n^0= a_p$.
With this input we obtain $A_{Kd}=-1.54 +i\, 1.29$ fm, much
closer to  the result of Ref.~\cite{torres}, $A_{Kd}=-1.34 +i 1.04$ fm.  
The remaining differences would give an idea of the accuracy of the FCA,
which seems to overpredict the Fadeev result by 15\%.
Note that the imaginary part has been much reduced when using the
scattering
lengths from Ref.~\cite{deloff}. 
This discussion shows clearly that the
basic reason for our large $K^-$ deuteron scattering length compared to
previous calculations lies on the different elementary amplitudes used,
not on the FCA approach. In particular, one can see from Table~\ref{tab:1} 
an especially large difference in the imaginary part of the
charge exchange amplitude $a_x$.
This explains why the relatively small differences observed for the $K^-$
deuteron scattering length when charge exchange effects are omitted (see
the ``only el.resc." results in Table~\ref{tab:2}) get enhanced in the
full calculation.

     A complement to the former discussion about the accuracy of the FCA
 approach can be drawn from the analysis performed in Ref.~ \cite{deloff}.
If one compares the FCA-integ and Faddeev results quoted there in
Table II it looks like the FCA-integ approach, which we have followed,
underestimates rather than overestimates
 the Faddeev results. 
However, the recoil factors have been omitted, as we have checked  by
reproducing
the single-channel FCA-integ results of that table setting $b_x=0$ in
Eq. (24) and the recoil factors $M_d/(m_K+M_d)$ and $(1+m_K/m_N)$ to 1.
When these factors are included the single-channel FCA-integ
result becomes $A_{Kd}=-0.997 +i\, 1.21$ fm, which overestimates by
15\% the
single-channel Faddeev result quoted there, $A_{Kd}=-0.85 +i\, 1.10$
fm, in agreement with what we have found for the
multichannel calculation.
We thus assume these uncertainties in our results, which can be further
improved by decreasing them in about 15\%, but we find this unnecessary
given the fact that uncertainties from other sources are as large as
that.
Once again, the main conclusion from this discussion is that the large
numbers found for the deuteron scattering length are due to the novel
elementary scattering amplitudes obtained in the chiral approach, and
particularly to the large contribution from the charge exchange terms.

\section{Conclusion}

 We have studied $K^-$ scattering on the deuteron at low energies and
have evaluated the $K^-d$ scattering length. The input consisted
on elementary ${\bar K} N$ amplitudes previously calculated using
chiral Lagrangians and a coupled channels unitary scheme.
We found that the multiple scattering series on the deuteron was
poorly convergent which forced us to sum it by means of a suitable
approximation to the Faddeev
equations. We found that we needed to include the charge exchange
channels in this approximate Faddeev approach, but we could omit intermediate
inelastic channels (involving for instance $\Sigma \pi$ states)
which, however, were relevant in the evaluation of the elementary
scattering matrices in Ref.~\cite{knangels}.

 We have found a $K^-d$ scattering length of the order $-1.62 +i\,1.91$
fm which has somewhat larger strength, both in the real and
imaginary parts, than in other approaches. We proved that these discrepancies
were mostly due to the novel elementary amplitudes from the chiral approach 
which were used in the calculation. We also found  that
isospin is only an approximate symmetry for $K^-d$ scattering
and violation of the isospin symmetry can be as large as 30\%,
hence, one should not rely upon isospin considerations when
evaluating the $K^-d$ scattering length.

 Comparison of the present results with the experimental results
expected  from the DEAR experiment at Frascati should bring light
on some of the issues involved in the problem, like chiral
symmetry and partial isospin breakup.

The findings of this paper should also be of much of use when
trying to extract information on elementary amplitudes from the
deuteron data. The formulas which we obtain would allow one to
deduce $a_n$ from $A_{Kd}$ using isospin relationships, but, as
discussed above, this would induce uncertainties of up to 30\%. We
have seen that the general formulas, without assuming isospin
symmetry, rely upon four scattering lengths
$a_p,\,a_n,\,a_x,\,a_n^0$. Knowledge of three of them from
other experiments and the use of the deuteron data would allow one
to obtain information on the fourth. Conversely, we can say that
the deuteron data will introduce a further check of consistency
between elementary amplitudes determined either experimentally or
theoretically.

\acknowledgments We would also like to acknowledge financial
support from the DGICYT under contracts PB96-0753, PB98-1247 and
AEN97-1693, from the Generalitat de Catalunya under grant
SGR98-11
and from the EU TMR network Eurodaphne, contract no.
ERBFMRX-CT98-0169.  S.S.K. is grateful to the Department of
Theoretical Physics and Instituto the F\'{\i}sica Corpuscular of
the University of Valencia for the hospitality extended during his
visit.

\begin{table}[htbp]
\caption {\small $K^-N$ scattering lengths (in fm) in the physical
and isospin bases }

\begin{tabular}{c|cc|c}
  reactions    & Physical basis    &  Isospin basis
&  Isospin basis, Ref.[19] \\
\hline $a_p\,\, (K^-p\rightarrow K^-p)$
&$-0.789 + i\, 0.929$ & $-0.799 + i\, 1.144$ & $-0.623 + i\, 0.763$ \\
\hline $a_n\,\, (K^- n\rightarrow K^-n)$
&$ 0.574 + i\, 0.619$ & $ 0.536 + i\, 0.521$ & $ 0.322 + i\, 0.748$ \\
\hline $a_x\,\,(K^-p\rightarrow {\bar K}^0 n)$
&$ -1.099 + i\, 0.522$ & $ -1.335 + i\, 0.623$ & $ -0.945 + i\, 0.015$
\\
\hline $a_n^0\,\,({\bar K}^0 n\rightarrow {\bar K}^0 n)$
&$ -0.387 + i\, 1.159$ & $ -0.799 + i\, 1.144$ & $ -0.623 + i\, 0.763$ \\
\end{tabular} 
\label{tab:1}
\end{table}

\begin{table}[htbp]
\caption {\small $K^-$-deuteron scattering length (in fm)
calculated using different approximations }

\begin{tabular}{c|cc|c}
approximations & Physical basis    & Isospin basis
&  Isospin basis, Ref.[19]  \\
\hline IA
& $ -0.260  + i\, 1.872 $ & $-0.318 + i\, 2.013 $ & $-0.364 + i\, 1.826 $
\\
\hline IA + double resc.
& $ -2.735  + i\, 2.895 $ & $-3.168 + i\, 3.717 $ & $-2.380 + i\, 1.485 $
\\
\hline IA + double+triple resc.
& $ -3.849  + i\, 2.963 $ & $-5.195 + i\, 4.935 $ & $-2.858 + i\, 0.089 $
\\
\hline $A_{Kd}$ (only el.resc.)
& $  -1.161 + i\, 1.336 $ & $-1.255 + i\, 1.518 $ & $-0.997 + i\, 1.212 $
\\
$A_{Kd}$ (charge exch.)
& $  -0.454 + i\, 0.573 $ & $-0.654 + i\, 0.937 $ & $-0.539 + i\, 0.079 $
\\
$A_{Kd}$ (total)
& $  -1.615 + i\, 1.909 $ & $-1.909 + i\, 2.455 $ & $-1.536 + i\, 1.291 $
\\
\end{tabular}
\label{tab:2}
\end{table}

\newpage

\begin{figure}[h]
\centerline{\psfig{file=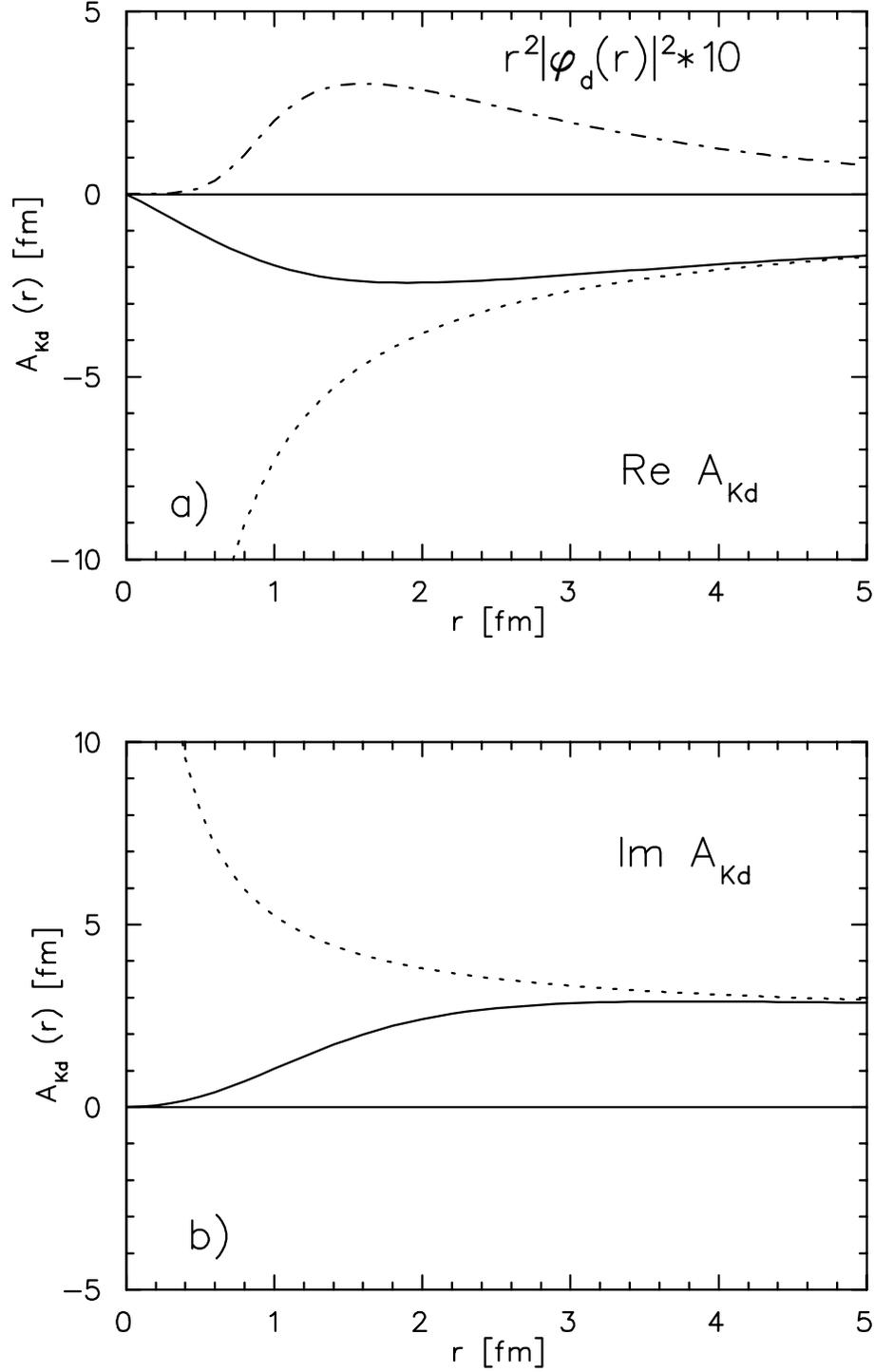,width=12cm,silent=}}
\vspace{0.5cm} \caption{ ${\hat A}_{Kd}(r)$ (solid curves) and
${\hat A}_{Kd}^{(1)}(r)$ (dotted curves). In the upper panel we
also show the deuteron wave function} \label{fig4_kor}
\end{figure}

\end{document}